\documentclass{article}

\usepackage{PRIMEarxiv}

\usepackage[utf8]{inputenc} 
\usepackage[T1]{fontenc}    
\usepackage{hyperref}       
\usepackage{url}            
\usepackage{booktabs}       
\usepackage{amsfonts}       
\usepackage{nicefrac}       
\usepackage{microtype}      
\usepackage{lipsum}
\usepackage{graphicx}
\graphicspath{{media/}}     

\title{Architecture-Aware Multi-Design Generation for Repository-Level Feature Addition}

\author{
  Mingwei Liu\\
  Sun Yat-sen University \\
  Zhuhai \\
  China\\
  \texttt{liumw26@mail.sysu.edu.cn}\\
   \And
  Zhenxi Chen\\
  Sun Yat-sen University \\
  Zhuhai \\
  China\\
  \texttt{chenzhx236@mail2.sysu.edu.cn} \\
  \And
  Zheng Pei\\
  Sun Yat-sen University \\
  Zhuhai \\
  China\\
  \texttt{peizh3@mail2.sysu.edu.cn} \\
  \And
   Zihao Wang  \\
  Sun Yat-sen University \\
  Zhuhai \\
  China\\
  \texttt{wangzh778@mail2.sysu.edu.cn} \\
  \And
  Yanlin Wang* \\
  Sun Yat-sen University \\
  Zhuhai \\
  China\\
  \texttt{yanlin-wang@outlook.com} \\
  \And
  Zibin Zheng \\
  Sun Yat-sen University \\
  Zhuhai \\
  China\\
  \texttt{zhzibin@mail.sysu.edu.cn}\\
}

\usepackage{float}
\usepackage{array}
\usepackage{rotating}  
\usepackage{pdflscape}
\usepackage{tabularx, multirow, makecell}

\usepackage{adjustbox}
\usepackage{amsmath,amssymb,amsfonts}
\usepackage{algorithmic}
\usepackage[linesnumbered,ruled,vlined]{algorithm2e}
\usepackage{setspace}
\usepackage{graphicx}
\usepackage{textcomp}
\usepackage{xcolor}
\usepackage{multirow}
\usepackage{mdframed}
\usepackage{balance}
\usepackage{booktabs}
\usepackage{threeparttable}
\usepackage{colortbl}
\usepackage{subfigure}
\usepackage{hhline}
\usepackage{pifont}
\usepackage{listings}
\usepackage{enumitem}
\usepackage{geometry}
\usepackage{verbatim}
\usepackage{xspace}
\usepackage{listings}
\newcolumntype{C}{>{\centering\arraybackslash}X}
\usepackage{pgfplots}
\pgfplotsset{compat=1.16}
\usetikzlibrary{pgfplots.statistics}

\newcommand{\rev}[1]{{#1}}

\definecolor{ggray}{HTML}{eff0f0}
\definecolor{gggray}{HTML}{E8E8E8}
\definecolor{ggggray}{HTML}{BEBEBE}

\newcommand{\ie}{\textit{i.e.,}\xspace}
\newcommand{\eg}{\textit{e.g.,}\xspace}

\newcommand{\drop}[1]{\textcolor{red!80!black}{\footnotesize ($\downarrow$#1\%)}}

\newcommand{\rise}[1]{\textcolor{teal}{\footnotesize ($\uparrow$#1\%)}}

\definecolor{LighterBlue}{RGB}{235, 245, 255}

\DeclareRobustCommand{\cnumcap}[1]{%
\textcircled{\scriptsize #1}%
}

\newcommand{\MyMethod}{RAIM}

\definecolor{myyellow}{HTML}{FFF2CC}

\newcounter{finding}
\newcommand{\finding}[1]{\refstepcounter{finding}
 	\vspace{1mm}
	\begin{mdframed}[linecolor=gray!25,roundcorner=12pt,backgroundcolor=myyellow!30,linewidth=3pt,innerleftmargin=2pt, leftmargin=0cm,rightmargin=0cm,topline=false,bottomline=false,rightline = false]
		\textbf{Finding \arabic{finding}:} #1
	\end{mdframed}
	\vspace{1mm}
}

\usepackage{tikz}

\begin{document}
\maketitle

\begin{abstract}

Implementing new features across an entire codebase presents a formidable challenge for Large Language Models (LLMs). This proactive task requires a deep understanding of the global system architecture to prevent unintended disruptions to legacy functionalities. Conventional pipeline and agentic frameworks often fall short in this area because they suffer from architectural blindness and rely on greedy single-path code generation. To overcome these limitations, we propose \MyMethod{}, a multi-design and architecture-aware framework for repository-level feature addition. This framework introduces a localization mechanism that conducts multi-round explorations over a repository-scale code graph to accurately pinpoint dispersed cross-file modification targets. Crucially, \MyMethod{} shifts away from linear patching by generating multiple diverse implementation designs. The system then employs a rigorous impact-aware selection process based on static and dynamic analysis to choose the most architecturally sound patch and avoid system regressions. Comprehensive experiments on the NoCode-bench Verified dataset demonstrate that \MyMethod{} establishes a new state-of-the-art performance with a 39.47\% success rate, achieving a 36.34\% relative improvement over the strongest baseline. Furthermore, the approach exhibits robust generalization across various foundation models and empowers open-weight models like DeepSeek-v3.2 to surpass baseline systems powered by leading proprietary models. Detailed ablation studies confirm that the multi-design generation and impact validation modules are critical to effectively managing complex dependencies and reducing code errors. These findings highlight the vital role of structural awareness in automated software evolution.
\end{abstract}


\section{Introduction}

The paradigm of software engineering is undergoing a transformative shift towards \textbf{natural language-driven feature addition}, a workflow that allows users to specify desired software behaviors through natural language descriptions rather than direct source code editing~\cite{hirzel2023low,rao2024ai}. Unlike traditional code generation tasks that predominantly focus on isolated function-level synthesis~\cite{yin2017syntactic,chen2021evaluating}, this task presents a complex challenge at the repository level. The recently introduced NoCode-bench~\cite{deng2025nocode} formalizes this critical software evolution scenario. In this task, the system takes \textbf{user-facing documentation changes}, such as release notes or updated specification documents, as input and is required to autonomously generate a corresponding code patch that implements the new features within an existing codebase. This task is paramount as it simulates real-world development workflows where documentation serves as the authoritative specification. It demands that LLMs not only interpret high-level functional requirements but also possess comprehensive architectural awareness to handle cross-file dependencies and ensure the new features seamlessly integrate without disrupting existing functionalities.

However, realizing this repository-level feature addition presents challenges that fundamentally differ from those in widely studied issue resolution tasks such as SWE-bench~\cite{jimenez2023swe}. While both fall under the umbrella of software maintenance, research indicates that approximately 60\% of maintenance costs are allocated to feature addition rather than mere defect correction~\cite{glass2001frequently,rahman2019supporting}. Existing methods developed for SWE-bench primarily focus on reactive maintenance, where the objective is to correct a specific deviation from expected behavior constrained by existing failing test cases. These approaches typically rely on fault localization techniques to identify a confined scope for repair. \textbf{In contrast, feature addition is a proactive architectural challenge requiring the implementation of high-level specifications without explicit fault signals.} This task requires agents not only to identify optimal insertion points among thousands of files but also to ensure that new additions integrate harmoniously with legacy modules, thereby avoiding regressions in complex, interdependent systems.

Although recent advancements in automated software engineering have demonstrated promising capabilities in complex repository maintenance, adapting these strategies to feature addition remains problematic. Representative state-of-the-art approaches typically fall into two categories: workflow-based methods like Agentless~\cite{xia2024agentless}, which employs a fixed multi-phase pipeline relying on hierarchical retrieval to locate edit positions and generate patches, and agent-based frameworks like OpenHands~\cite{wang2024openhands}, which utilize an iterative loop of thought and tool execution to navigate the codebase. \textbf{While effective for issue resolution, these approaches exhibit critical limitations in architectural decision-making due to their lack of repository-level architectural awareness and linear implementation strategies that bypass rigorous impact assessment.} First, by treating repositories largely as unstructured text during retrieval, they lack a holistic view of cross-file structural dependencies~\cite{ouyang2024repograph,ruan2024specrover}, often missing the optimal insertion points for new logic. Second, both paradigms tend to adopt a linear or greedy generation strategy that blindly accepts the first plausible patch. This lack of rigorous evaluation is particularly detrimental in feature addition, where the primary risk is not just functionality failure but the subtle disruption of existing architectural patterns and downstream dependencies.

To address these limitations and enhance LLMs' capabilities in feature addition development scenarios, we propose \textbf{\MyMethod{}}, a \textbf{R}epository-level \textbf{A}rchitecture-aware feature \textbf{I}mplementation framework based on \textbf{M}ulti-design. First, to overcome the lack of architectural awareness, \MyMethod{} constructs a repository-level code graph that captures semantic and structural relationships including function calls and class inheritance. By performing multi-round iterative searches on this graph, our approach can trace the implementation patterns of similar existing features and guide the localization to the architecturally correct modules. Second, to avoid local optimal in code generation, \MyMethod{} employs a multi-design generation strategy. Instead of generating a single patch, the system prompts the LLM to brainstorm diverse implementation strategies which expands the solution space. Finally, to ensure the robustness required for production-grade software, a patch selection mechanism based on \textbf{code change impact analysis} is introduced, which evaluates the influence of proposed modifications on both the stability of existing functionalities and the correctness of the new feature. By statically analyzing the scope of code changes and dynamically executing tests, \MyMethod{} rigorously evaluates the potential risks of each candidate patch and selects the patch that best balances functional correctness with architectural consistency.

To evaluate the effectiveness of \MyMethod{}, we evaluate it on the NoCode-bench Verified dataset~\cite{deng2025nocode}, a benchmark for repository-level feature addition. Experimental results show that \MyMethod{} consistently outperforms state-of-the-art baselines. In particular, when instantiated with the Gemini-2.5-Pro model, \MyMethod{} achieves a success rate of 39.47\%, surpassing the previous best-performing method, Agentless, by a substantial margin 36.34\%. Furthermore, \MyMethod{} enables the open-source DeepSeek-v3.2 to achieve a 34.21\% success rate, outperforming baseline methods utilizing powerful closed-source models.  Further analysis demonstrates that \MyMethod{} maintains robust performance across various models, yielding consistent performance gains ranging from 9.7\% to 221.4\% and specifically achieving a 191.7\% relative improvement in complex cross-file modification tasks. Beyond the overall performance comparison, we further analyze \MyMethod{} from multiple perspectives, including its generalization across 7 LLMs, its effectiveness on cross-file feature addition tasks, and the contributions of its key design components.

In summary, this paper makes the following contributions:
\begin{itemize}[itemsep=2pt, topsep=2pt, parsep=0pt, partopsep=0pt]
    \item An \textbf{architecture-aware feature localization strategy} based on repository-level code graph search is proposed, employing multi-round iterative exploration to accurately identify cross-file editing locations for feature addition.
    \item An \textbf{impact-aware multi-design patch generation framework} is introduced to expand the solution space through diverse implementation design and to select high-quality patches via change impact analysis, ensuring architectural consistency and preventing regressions.
    \item \textbf{Extensive experimental evaluation} is conducted on NoCode-bench Verified, demonstrating that \MyMethod{} achieves state-of-the-art performance for repository-level feature addition.
\end{itemize}

\section{Motivation Example}

Repository-level feature addition is more architecturally demanding than issue resolution, requiring seamless integration of new logic without disrupting legacy functionality. To illustrate these challenges, we analyze \texttt{pylint-dev\_\_pylint-8190} from NoCode-bench\footnote{\url{https://huggingface.co/NoCode-bench}}, which involves adding a \texttt{-{}-show-stdlib} option to Pylint's UML generator. This task requires following the design pattern of the existing \texttt{-{}-show-builtin} option to control module visibility in generated diagrams.
\begin{figure}[htbp]
  \centering
  \includegraphics[width=1.0\linewidth]{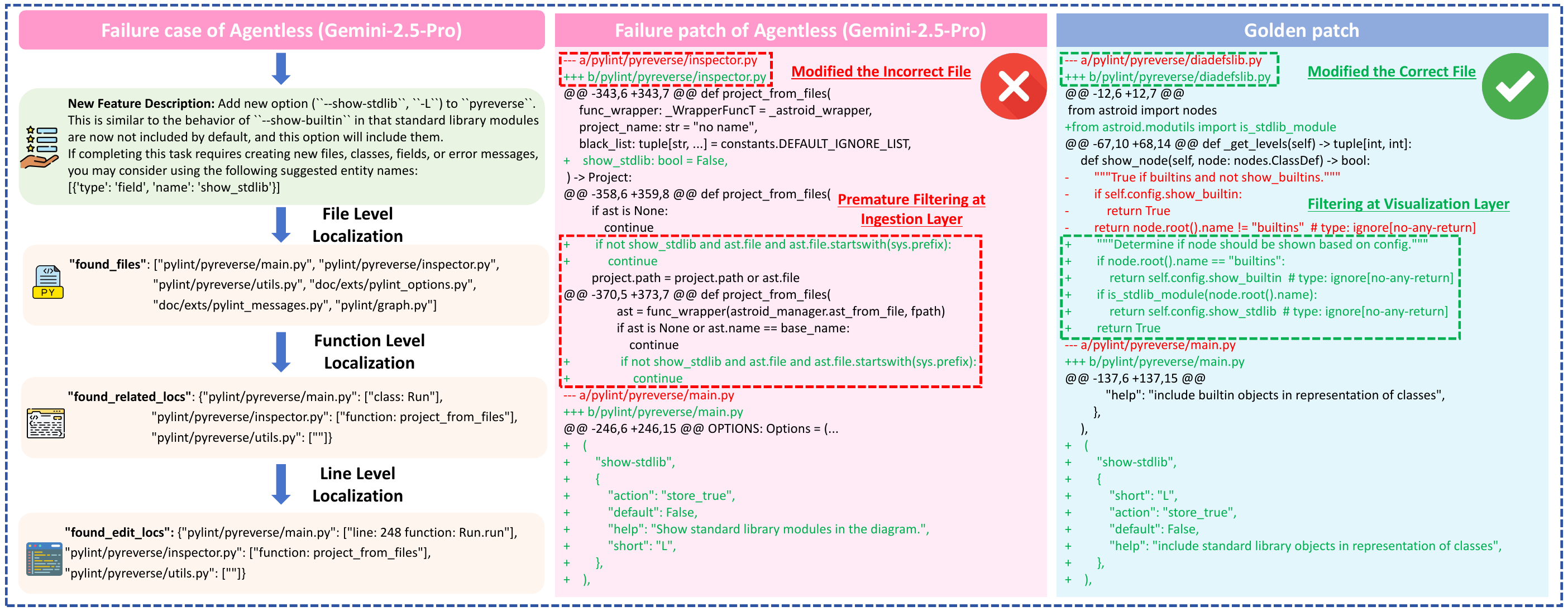} 
  \caption{Motivating example on instance \texttt{pylint-dev\_\_pylint-8190}.}
  \label{fig:motivation}
\end{figure}

Agentless~\cite{xia2024agentless} relies on hierarchical retrieval for localization and stochastic sampling for generation. This linear workflow lacks multi-design exploration and fails to perform systematic impact analysis during patch selection. Such deficiencies lead to critical failures in feature addition, where successful implementation requires a holistic evaluation of code changes rather than satisfying localized semantic similarity.

\textbf{(1) Lack of Architectural Awareness in Cross-File Contexts.} Agentless demonstrates architectural blindness by targeting \texttt{inspector.py} (ingestion layer) instead of the visualization-layer file \texttt{diadefslib.py}. While \texttt{inspector.py} handles file parsing, filtering standard library modules at this stage prematurely discards essential AST nodes, breaking inheritance analysis downstream. In contrast, the correct implementation suppresses these modules only during rendering to preserve structural data. By treating the repository as unstructured text, the baseline overlooks systemic layers, resulting in an architecturally inappropriate modification.

\textbf{(2) Lack of Impact Analysis.} Agentless fails to evaluate the repercussions of code modifications on overall system stability. While the baseline literally fulfills the requirement by filtering modules in \texttt{inspector.py}, this implementation introduces critical regressions by discarding AST nodes during the parsing stage. Consequently, parent node information for user-defined classes inheriting from the standard library is lost, which breaks inheritance relationships in the final output. Because Agentless lacks a mechanism to analyze code change impacts and relies on stochastic sampling without multi-design exploration, it accepts plausible implementation that compromises structural integrity. This case emphasizes the necessity of impact verification to ensure that feature completion does not degrade the existing codebase.

\textbf{Key Insights.} These failure modes suggest that a robust framework must move beyond localized editing. An effective approach should: first, demonstrate \textbf{repository-level architectural awareness} to identify modification points that align with internal system layers; second, incorporate \textbf{systematic code change impact analysis} to assess repercussions and prevent regressions. By integrating architectural navigation with impact-driven implementation selection, a system can ensure that new features are integrated seamlessly while preserving codebase integrity.
\section{Approach}

\begin{figure}[htbp]
  \centering
  \includegraphics[width=1.0\linewidth]{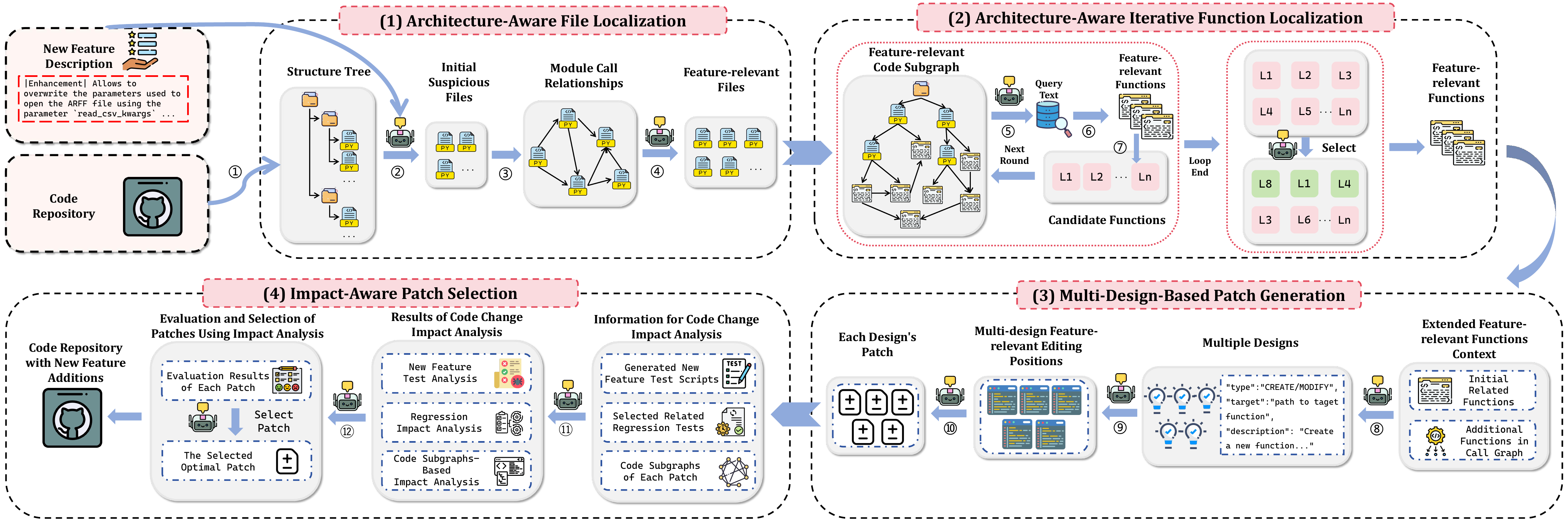}
  \caption{Overview of \MyMethod{}. \textbf{\cnumcap{1}} Parse the code repository into a structure tree; 
 \textbf{\cnumcap{2}} Localize initial suspicious files; 
 \textbf{\cnumcap{3}} Analyze module call relationships of suspicious files; 
  \textbf{\cnumcap{4}} Localize feature-relevant files; 
  \textbf{\cnumcap{5}} Generate query text for feature-relevant functions; 
  \textbf{\cnumcap{6}} Retrieve feature-relevant functions from neighbor nodes; 
  \textbf{\cnumcap{7}} Update candidate functions with retrieved results; 
  \textbf{\cnumcap{8}} Generate multiple feature-relevant designs; 
  \textbf{\cnumcap{9}} Localize line-level editing positions; 
  \textbf{\cnumcap{10}} Generate diverse patches based on multi-design; 
  \textbf{\cnumcap{11}} Perform code change impact analysis; 
  \textbf{\cnumcap{12}} Evaluate patches based on impact analysis results. }
  \label{fig:overview}
\end{figure}

Motivated by the challenges and insights discussed in the previous section, we present \MyMethod{}, a repository-level architecture-aware framework for feature addition. Fig.~\ref{fig:overview} provides an overview of the framework. \MyMethod{} takes a feature description and the repository’s structure tree as input and outputs an optimal patch for the new feature. The framework consists of four main stages: architecture-aware file localization, architecture-aware iterative function localization, multi-design patch generation, and impact-aware patch selection. 

At the file-level localization stage, feature-relevant files are identified using repository-level call relationships, and the top-ranked files are selected as anchors for function-level localization (Section~\ref{file-level}). During function-level localization, multi-round search over feature-relevant code subgraphs adds functions to the candidate function set, which is reranked iteratively to produce the final set of feature-relevant functions (Section~\ref{func-level}). In the patch generation stage, diverse patches are produced through multiple distinct implementation designs, exploring different editing locations (Section~\ref{planning}). Finally, candidate patches are evaluated using code change impact analysis, and the optimal patch is selected for implementation (Section~\ref{patch selection}).

\begin{figure}[htbp]
  \centering
  \includegraphics[width=\linewidth]{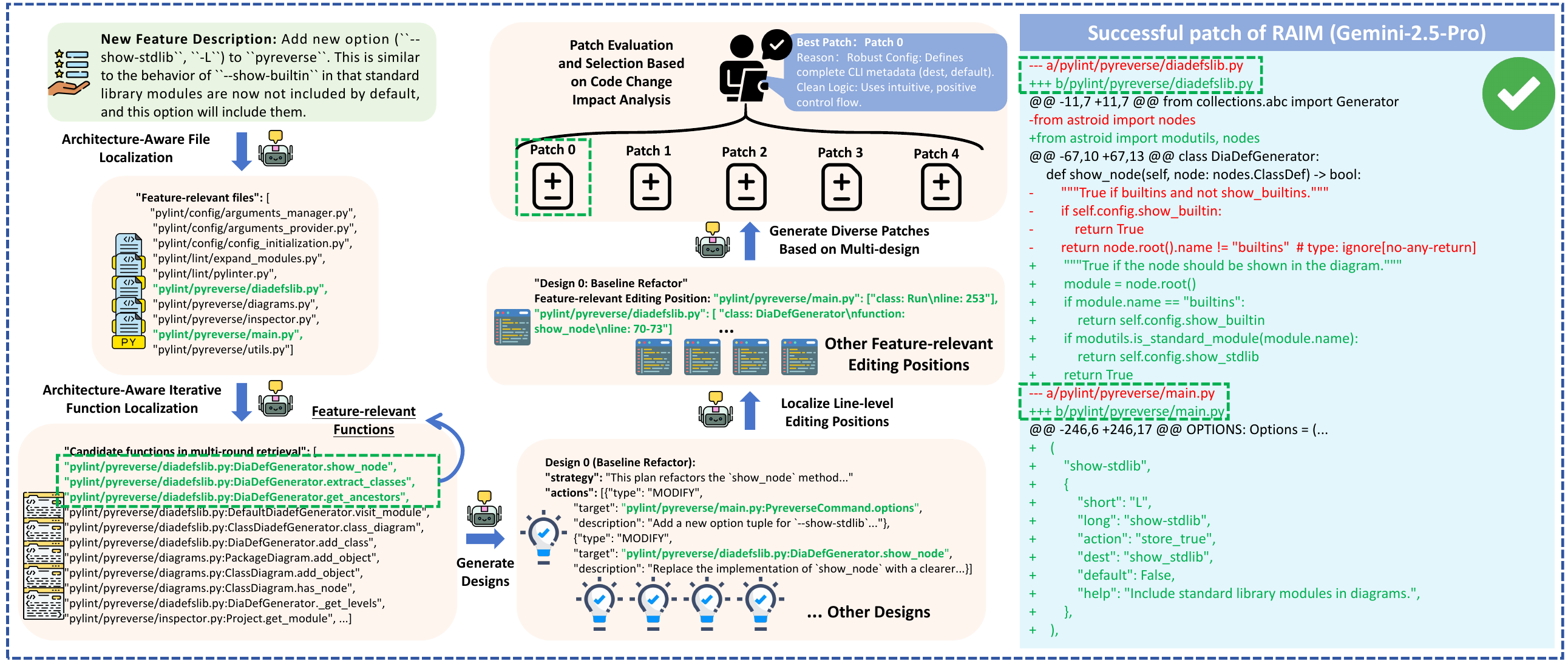}
  \caption{The example of \MyMethod{} on instance \texttt{pylint-dev\_\_pylint-8190}.}
  \label{fig:example}
\end{figure}

\subsection{Architecture-Aware File Localization}
\label{file-level}
To address the challenge of identifying scattered modification points across interdependent modules in repository-level feature addition, an architecture-aware file localization strategy is proposed. This phase takes the new feature description and the code repository as input, producing a ranked list of feature-relevant files.
Given that directly processing the entire codebase for LLMs is impractical, the repository is first parsed into a hierarchical structure tree. The LLM utilizes this tree combined with the feature description to preliminarily identify a set of initial suspicious files.
Subsequently, to capture the architectural dependencies required for implementing new features, import relationships within these files are analyzed to construct a file-level call graph. Finally, the repository structure tree, file-level call graph, and the code skeleton of the initial suspicious file are input into the LLM to localize feature-relevant files.

As shown in Fig.~\ref{fig:example}, for the \texttt{pylint-dev\_\_pylint-8190} instance, this strategy successfully retrieves a comprehensive list of feature-relevant files, most notably including \texttt{pylint/pyreverse/diadefslib.py}, which is critically missed by the baseline method (Agentless). While Agentless restricts its scope to ingestion-related files like \texttt{inspector.py}, the analysis of call relationships in \MyMethod{} reveals the architectural dependency between the configuration entry in \texttt{main.py} and the visualization logic in \texttt{diadefslib.py}. This ensures that the correct target file is included for the subsequent fine-grained localization.


\subsection{Architecture-Aware Iterative Function Localization}
\label{func-level}
\begin{figure}[htbp]
  \centering
  \includegraphics[width=0.8\linewidth]{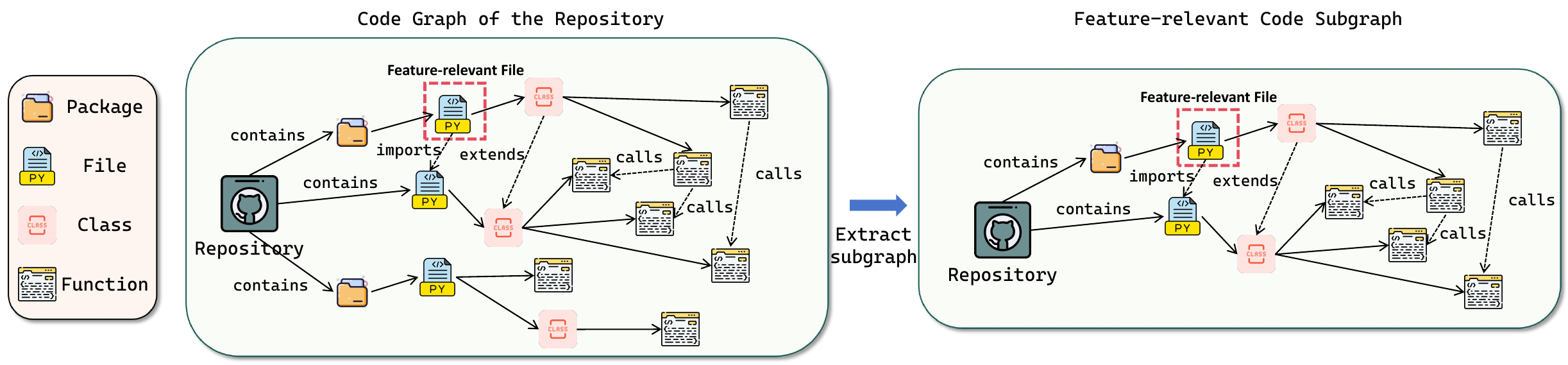}
  \caption{Construction of Feature-relevant Code Subgraph.}
  \label{fig:code_graph}
\end{figure}

After obtaining feature-relevant files, a directed code graph $\mathcal{G} = (\mathcal{V}, \mathcal{E})$ (\ie Feature-relevant Code Subgraph) is constructed to capture repository-level dependencies. As illustrated in Fig.~\ref{fig:code_graph}, the set of nodes $\mathcal{V}$ consists of code entities at varying granularities, including \texttt{Package}, \texttt{File}, \texttt{Class}, and \texttt{Function}. The edges $\mathcal{E}$ capture both hierarchical and semantic dependencies: \textit{contains} edges represent file system structure and code encapsulation, while \textit{imports}, \textit{extends}, and \textit{calls} edges represent reference relationships derived via static AST analysis.
To construct the feature-relevant search space, the identified feature-relevant files are mapped to corresponding \texttt{File} nodes in $\mathcal{G}$ to serve as anchor nodes. Subsequently, a Feature-relevant Code Subgraph is extracted by expanding these anchors to include their one-hop neighbor nodes. This process generates a compact subgraph that preserves both the target code entities and their direct structural dependencies. Using this subgraph, \MyMethod{} performs multi-round function localization, which primarily consists of two steps.

\textbf{Step I: Embedding model-based multi-round function retrieval}
To navigate the code subgraph and identify semantically related functions that keyword matching might miss, a multi-round retrieval strategy is employed. This step takes the \textbf{feature-relevant code subgraph} and the \textbf{new feature description} as input, iteratively populating a set of candidate functions.

First, embedding model, specifically Qwen3-Embedding's text-embedding-v4~\cite{zhang2025qwen3}, is utilized to convert the source code of \texttt{Function} nodes within the subgraph into dense vector representations.
In the first round, a dual-path approach is adopted: the LLM identifies the top-$k$ related functions based on the feature description and file skeletons, while the embedding model retrieves the top-$k$ functions via cosine similarity using the feature description as a query. Both sets are added to the candidate functions.
In subsequent rounds, the LLM analyzes the current candidates to determine if further retrieval is necessary. If affirmative, a new search query (\eg ``Find where the show-builtin option is defined and $\dotsc$'') is generated based on the context of existing candidate functions, vectorized, and used to search the neighbor nodes of previously located functions. The top-$n$ relevant neighbors are iteratively added to the candidate functions.
As shown in Fig.~\ref{fig:example}, the previous file localization phase produced a broad list of feature-relevant files, containing several interference items such as \texttt{pylint/config/arguments\_manager.py}. However, the multi-round retrieval process effectively narrows this scope. By traversing the code graph, the generated set of candidate functions heavily concentrates on the logic within \texttt{pylint/pyreverse/diadefslib.py}. This demonstrates that the retrieval mechanism filters out the noise from the broad search space and accurately zooms in on the code regions most critical for the new feature's implementation.

\textbf{Step II: LLM-based feature-relevant function selection}
To filter noise from the accumulated candidates and pinpoint the precise editing locations, an LLM-based reranking strategy is implemented. This step accepts the candidate functions as input and outputs the top-$m$ most relevant functions.
The new feature description and the source code of the functions in the candidate set are fed into the LLM for relevance assessment. To accommodate context window constraints, each function is truncated to a predefined token limit from the end. The LLM then reranks these candidates based on their semantic alignment with the feature requirements.
As illustrated in Fig.~\ref{fig:example}, the set of candidate functions obtained from the previous step still contains numerous neighbor nodes (\eg \texttt{DiaDefGenerator.extract\_classes}) and residual noise from other modules (\eg \texttt{Project.get\_module} in \texttt{inspector.py}). The selection strategy leverages the LLM's semantic understanding to discern the specific utility of each candidate. Consequently, \texttt{DiaDefGenerator.show\_node} is correctly identified as the top-ranked function, distinguishing the precise editing location from the surrounding context and interference.

\subsection{Multi-Design-Based Patch Generation}
\label{planning}
To avoid the limitations of generating a single fragile edit and to mimic the rigorous engineering process of brainstorming diverse strategies, a multi-design-based patch generation phase is proposed. This phase expands the solution space by instructing the model to brainstorm diverse implementation plans before executing code modifications. The process consists of two steps: multi-design generation and multi-design-based patch generation.

\textbf{Step I: Multi-design generation}
To bridge the gap between local function-level edits and system-wide architectural consistency, a multi-design generation strategy powered by call-graph contextual enrichment is proposed. This module takes the feature-relevant functions, the code graph, and the feature description as input, producing $N$ distinct architectural designs.
In this phase, the implementation process is modeled after a professional software architect who first assesses the broader system impact before drafting a blueprint. Specifically, the call and dependency relationships of the previously identified feature-relevant functions are parsed from the code graph into a tree-structured representation. This structural context is then presented to the LLM to analyze potential co-modifications or downstream impacts. By identifying these additional function contexts, the model gains a holistic view of the repository, enabling it to brainstorm $N$ distinct designs from various perspectives, such as refactoring or pattern extension. Each design consists of a sequence of actions, where each action explicitly comprises: (1) a \texttt{type} defining the kind of modification (\eg \texttt{MODIFY} or \texttt{CREATE}); (2) a \texttt{target} specifying the function or file path to be modified; and (3) a \texttt{description} providing a high-level natural language explanation of the intended change logic.
As illustrated in Fig.~\ref{fig:example}, this step is crucial for cross-file implementation. While the initial function-level localization focuses primarily on the core logic in \texttt{diadefslib.py}, the extraction of additional context through call-graph analysis reveals the necessity of a coordinated modification in \texttt{main.py} to support the new command-line option. Consequently, Design 0 successfully incorporates this cross-file joint modification, ensuring that the final feature is architecturally complete and well-integrated into the entry point of the system.

\textbf{Step II: Multi-design-based patch generation}
To bridge the semantic gap between abstract architectural designs and executable code while preserving solution diversity, a \textbf{multi-design-based patch generation} mechanism is employed. This step takes the set of \textbf{architectural designs produced in the preceding step} and the \textbf{codebase} as input, outputting a corresponding set of diverse \textbf{code patches}.
Since different designs target distinct logic paths, specialized line-level localization is performed independently for each design. To facilitate this localization, an intelligent fallback and dynamic granularity strategy is employed to construct the input context for the model:
(1) For \texttt{MODIFY} actions, context is initially targeted at specific lines of existing entities.
(2) For \texttt{CREATE} actions, the context granularity is dynamically adjusted: when adding a method to an existing class, the relevant class code is provided; conversely, when creating new classes or functions, a file skeleton is extracted to provide a structural overview.
Based on this tailored context, the LLM accurately localizes the editing positions. Finally, guided by these precise editing positions, the LLM generates corresponding candidate patches, forming a diverse set $\mathcal{P} = \{P_0, P_1, ..., P_{N-1}\}$. Referring to Fig.~\ref{fig:example}, Patch 0 is generated based on Design 0, where the model accurately leverages the context of \texttt{main.py} and \texttt{diadefslib.py} to implement the feature. By generating diverse patches derived from multiple designs, a rich pool of candidates is provided for the subsequent impact analysis.

\subsection{Impact-Aware Patch Selection}
\label{patch selection}
In the previous stage, a set of candidate patches $\mathcal{P} = \{P_0, P_1, \dots, P_{N-1}\}$ is generated, with each patch representing a distinct implementation strategy. However, patches generated by LLMs may correctly implement new features while potentially disrupting existing system architectures. To address this, \MyMethod{} employs a rigorous screening mechanism that combines static code analysis with dynamic testing analysis to analyze code change impacts, thereby selecting an optimal patch.


\subsubsection{\textbf{Code change impact analysis}}

To rigorously evaluate the candidate patches $\mathcal{P}$, a comprehensive impact analysis mechanism is employed, combining static architectural assessment with dynamic behavioral verification. This process is divided into three distinct dimensions:

\textbf{Code Subgraph-Based Impact Analysis.}
To statically assess the structural repercussions of code modifications, a graph-driven analysis is performed utilizing Abstract Syntax Tree (AST) parsing and the repository-level code graph. Entities modified within a patch are first identified and mapped onto the code graph to expand a specific code subgraph, which isolates the dependency structure associated with the changes. The output of this step is a static change impact analysis report consisting of three core components. First, the report identifies high-impact nodes based on their in-degree and out-degree centrality within the subgraph to define their impact scope. For these nodes, the report provides critical architectural data, including the interface parameter list, the number of upstream call locations, and the count of downstream dependent nodes. To illustrate, for a critical entity such as \texttt{main.py:Run}, the report identifies an impact scope where the class is called by 12 upstream locations and depends on 7 downstream entities. Second, the report identifies the patched entities, specifically listing the modified functions and classes such as \texttt{show\_node} to define the precise modification scope. Third, the step expands the candidate patch context by integrating the generated changes with contiguous surrounding lines of code and their corresponding line numbers from the source repository.

\textbf{Regression Impact Analysis.}
To verify that a new feature does not compromise existing system stability, a regression impact analysis is conducted. First, nodes modified by the patch are identified. The code graph is then traversed from these nodes to pinpoint dependent test nodes, thereby selecting a suite of relevant test cases. To establish a reliable baseline, this suite is executed on the codebase prior to applying the patch. Only the test cases that initially pass are retained for the actual regression check. The patch is then applied, and this filtered set of tests is re-executed. Any test that transitions from a passing to a failing state signifies a regression. In such cases, an LLM is utilized to analyze the corresponding error logs and summarize the root causes of the failure.

\textbf{New Feature Test Analysis.}
To verify the functional correctness of the newly implemented feature, a dedicated test analysis is performed. Initially, an LLM is prompted to generate new test scripts tailored to the feature's requirements as described in the documentation changes. After the candidate patch is applied, these newly generated scripts are executed against the modified codebase. The resulting execution logs are then provided to an LLM for a comprehensive summary of the outcomes. In the event of test failures, the LLM is further tasked with analyzing the error logs to summarize the root causes of the failure. This process determines whether the patch correctly implements the intended functionality.

\subsubsection{\textbf{Impact-Driven Patch Evaluation and Selection}}
\label{selection}
To ensure the final submission is not only functionally correct but also robust and architecturally consistent, a rigorous screening mechanism is employed. This phase utilizes the insights derived from the code change impact analysis to evaluate and select the optimal patch. The process is divided into two stages: individual patch evaluation and comprehensive patch selection.

\textbf{Impact-aware patch evaluation.}
To assess the quality and risk of each generated candidate, a multi-dimensional evaluation mechanism is implemented, acting as an automated senior code reviewer. This module takes the \textbf{new feature description}, the \textbf{Code Subgraph--Based Impact Analysis}, the \textbf{Regression Impact Analysis}, and the \textbf{New Feature Test Analysis} as input, producing a structured assessment report for each patch.
The LLM evaluates each candidate's strengths and weaknesses across these three dimensions. It scrutinizes regression risks, validates functional correctness based on new feature tests, and assesses architectural integrity using the static impact analysis. Based on this holistic assessment, each patch is scored across five key criteria, with each criterion rated on a scale of 0 to 2: (1) \textbf{Relevance}: how well the code implements the core functionality; (2) \textbf{Syntax and Style}: whether the code is valid and well-structured; (3) \textbf{Upstream Safety}: whether it introduces breaking changes for upstream callers; (4) \textbf{Downstream Correctness}: whether downstream APIs are used correctly; and (5) \textbf{Regression Safety}: whether it maintains compatibility with existing functionality.
This scoring framework is designed to balance the functional correctness of the new feature implementation with the preservation of architectural integrity, ensuring that the generated patches successfully fulfill the requirements while respecting the structural constraints of the repository.

\textbf{Impact-aware optimal patch selection.}
To identify the optimal solution from the pool of evaluated candidates, a comprehensive selection strategy is employed. This step aggregates the evaluation reports of all patches, enabling a final decision based on a holistic assessment.
The selection process involves a unified evaluation where the LLM considers multiple critical aspects simultaneously. These include the impact on existing functionality to avoid regressions, the correctness of the new feature implementation to ensure requirements are met, and the overall code quality and architectural risk to promote maintainability. The model balances these factors to select the most well-rounded and robust solution.

As shown in Fig.~\ref{fig:example}, multiple patches are evaluated through this lens. Patch 0 is selected as the best patch because it demonstrates an optimal balance across all dimensions. It correctly implements the new feature while passing regression tests. Crucially, its code quality and architectural design, described as having Robust Config and Clean Logic, are superior. In contrast, other patches are rejected due to either failing regression checks or exhibiting higher architectural risk as identified by the impact analysis, such as modifying the incorrect architectural layer.


\section{Evaluation}
\label{sec:evaluation}
To evaluate \MyMethod{} for repository-level feature addition, we address five RQs moving from high-level performance and cross-file comparisons (RQ1–RQ2) to in-depth analysis of internal strategies (RQ3–RQ4), and a detailed failure analysis investigating its capability to overcome baseline limitations (RQ5).

\begin{itemize}
    \item \textbf{RQ1:} How does \MyMethod{} compare with state-of-the-art baselines in terms of effectiveness and generalizability for repository-level feature addition?

    \item \textbf{RQ2:} How well does \MyMethod{} handle cross-file feature additions?

    \item \textbf{RQ3:} What is the contribution of each component in \MyMethod{}?

    \item \textbf{RQ4:} How effective are the multi-design and patch selection strategies in \MyMethod{}?

    \item \textbf{RQ5:} How does \MyMethod{} address common failure cases in existing methods?
\end{itemize}

\subsection{Experiment Setup}
This section details the experimental setup, including the dataset, baselines, evaluation metrics, and implementation details.
\subsubsection{\textbf{Dataset}}
The evaluation is conducted on the NoCode-bench benchmark~\cite{deng2025nocode}, specifically its \textbf{Verified} subset, which comprises 114 high-quality, manually validated problems. Each instance in NoCode-bench formalizes a repository-level feature addition task. For each task, the benchmark provides the problem inputs, including a \textbf{documentation change} that specifies the new feature and a complete code repository at a specific pre-feature commit. The objective is to generate a code patch that correctly implements the specified feature. To assess the correctness of the generated patch, NoCode-bench supplies a test oracle composed of two distinct developer-written test suites: a set of \textbf{validation tests} (F2P) that are expected to transition from failing to passing, and a set of \textbf{regression tests} (P2P) that must remain passing to ensure no existing functionality is broken.

\subsubsection{\textbf{Baselines}}
We compare \MyMethod{} with two open-source approaches on the NoCode-bench benchmark, representing workflow-based (Agentless~\cite{xia2024agentless}) and agent-based (OpenHands~\cite{wang2024openhands}) paradigms for repository-level feature addition.

\textbf{Agentless} adopts a deterministic, multi-stage workflow that progressively narrows the search space via hierarchical retrieval. It first identifies relevant files, then localizes specific classes or functions, and finally pinpoints the exact code locations for modification. Patch generation is performed once sufficient context is collected. We use the official implementations Agentless-1.0~\cite{xia2024agentless} as the workflow-based baseline.

\textbf{OpenHands} represents an agent-based framework that performs feature addition through iterative planning, reasoning, and tool execution. An autonomous agent dynamically interacts with the repository by invoking tools such as file editors and shell commands, with intermediate results fed back to guide subsequent actions. In our experiments, we use the OpenHands CodeAct agent~\cite{wang2024openhands} in a sandboxed environment.


\subsubsection{\textbf{Metrics}}
To evaluate the performance of the proposed method on the NoCode-bench dataset, the evaluation metrics utilized are based on prior research \cite{deng2025nocode}: the success rate of feature addition tasks (Success(\%)), the regression tests pass rate (RT(\%)), and feature validation (FV).

\textbf{Success Rate (Success(\%)):} The success rate of feature addition tasks to evaluate the problem-solving capability of LLMs. This metric measures the percentage of issues where feature additions are successfully implemented to the total number of issues.

\textbf{Regression Tests Pass Rate (RT(\%)):} The percentage of tasks passing all regression tests after applying the generated patch.

    
\textbf{Feature Validation (FV):} Following NoCode-bench~\cite{deng2025nocode}, FV is defined as the micro-level and macro-level pass rates of F2P tests:
$\text{FV-Micro} = \frac{\sum_{i=1}^{N} \#Passed_i}{\sum_{i=1}^{N} \#Total_i}$,
$\text{FV-Macro} = \frac{1}{N} \sum_{i=1}^{N} \frac{\#Passed_i}{\#Total_i}$.


\subsubsection{\textbf{Implementations}}
The implementation of \MyMethod{} utilizes a diverse suite of state-of-the-art LLMs selected based on their advanced reasoning capabilities and established efficacy in software engineering tasks. The evaluation encompasses leading closed-source models, including \textbf{GPT-5-Chat} and \textbf{Gemini-2.5-Pro} (a reasoning model), alongside prominent open-source counterparts such as \textbf{DeepSeek-v3-0324}, \textbf{DeepSeek-R1-0528} (a reasoning model), \textbf{Qwen3-235B-A22B} (leveraging thinking mode), \textbf{DeepSeek-v3.2}, and \textbf{DeepSeek-v3.2-thinking}. All models are accessed through their respective official APIs. To ensure deterministic and reproducible outputs, the temperature parameter is set to 0 for all generation tasks. Within the framework, text-embedding-v4 model \cite{zhang2025qwen3} is employed to compute dense vector representations. Regarding the hyperparameters for the multi-round function localization, the initial retrieval is set to identify the top-$k$ ($k=3$) functions, the neighbor expansion considers the top-$n$ ($n=5$) related nodes, and the final selection retains the top-$m$ ($m=3$) most relevant candidates. For the multi-design generation phase, the system is configured to produce $N=5$ distinct implementation designs per task. Furthermore, to maintain the integrity of the autonomous evaluation as requested by the NoCode-bench maintainers, the pre-identified test lists within the \texttt{PASS2PASS} and \texttt{FAIL2PASS} fields are intentionally excluded from the selection and execution process to ensure a rigorous assessment of the system's independent implementation capabilities.

\subsection{RQ1: Effectiveness and Generalizability of \MyMethod{}}
\subsubsection{Design}
To evaluate the effectiveness of \MyMethod{}, it is compared against two baseline methods, Agentless and OpenHands, on the NoCode-bench Verified dataset. To further analyze the generalization of \MyMethod{} across different LLMs, it is compared against the superior method Agentless on seven LLMs using the NoCode-bench Verified dataset.

\begin{table}[htbp]
\footnotesize
\renewcommand{\arraystretch}{0.8} 
\setlength{\tabcolsep}{5pt}       
  \centering
  \caption{Performance of methods on NoCode-bench Verified. Results for OpenHands and Agentless from~\cite{deng2025nocode}.}
  \label{tab:ncb_verified_results}
  
  \begin{tabular}{ll ccc >{\columncolor{LighterBlue}}c} 
    \toprule
    \textbf{Method} & \textbf{Model} & \textbf{RT (\%)} & \textbf{FV-Micro (\%)} & \textbf{FV-Macro (\%)} & \textbf{Success (\%)} \\
    \midrule
    \multirow{5}{*}{OpenHands} 
        & Qwen3-235B             & 47.37 & 1.96  & 14.03 & 7.89  \\
        & DeepSeek-R1            & 46.49 & 0.47  & 10.86 & 7.02  \\
        & DeepSeek-v3            & 49.12 & 1.68  & 18.29 & 11.40 \\
        & Gemini-2.5-Pro         & 61.40 & 0.01  & 0.29  & 0.00  \\
        & Claude-4-Sonnet        & 69.30 & 11.25 & 36.48 & 25.44 \\
    \midrule
    \multirow{8}{*}{Agentless} 
        & Qwen3-235B             & 76.32 & 8.75  & 22.39 & 13.16 \\
        & GPT-5-Chat             & 82.46 & 8.50  & 33.01 & 18.42 \\
        & DeepSeek-R1            & 73.68 & 10.87 & 35.52 & 25.44 \\
        & DeepSeek-v3            & 78.95 & 7.96  & 32.80 & 21.05 \\
        & DeepSeek-v3.2          & 28.95 & 9.46  & 37.42 & 28.95 \\
        & DeepSeek-v3.2-thinking & 79.82 & 8.41  & 37.02 & 27.19 \\
        & Gemini-2.5-Pro         & 74.56 & 6.22  & 20.55 & 12.28 \\
        & Claude-4-Sonnet        & 79.82 & 8.47  & 38.48 & 28.07 \\
    \midrule
    \multirow{7}{*}{\MyMethod{}} 
        & Qwen3-235B             & 79.82          & 9.76           & 27.45          & 16.67          \\
        & GPT-5-Chat             & \textbf{89.47} & 13.43          & 32.33          & 21.93          \\

        & DeepSeek-v3            & 81.58          & 15.14          & 35.64          & 25.44          \\
        & DeepSeek-R1            & 77.19          & 12.47          & 41.79          & 29.82          \\
        & DeepSeek-v3.2          & 85.96          & 16.01          & 45.58          & 34.21          \\
        & DeepSeek-v3.2-thinking & 78.07          & 11.93          & 41.74          & 29.82          \\
        & Gemini-2.5-Pro         & 82.46          & \textbf{17.16} & \textbf{52.09} & \textbf{39.47} \\
    \bottomrule
  \end{tabular}
\end{table}

\subsubsection{Results}
Table \ref{tab:ncb_verified_results} shows the performance comparison of \MyMethod{} against existing baseline models on the NoCode-bench Verified dataset.
\begin{figure}[htbp]
  \centering
  \includegraphics[width=0.6\linewidth]{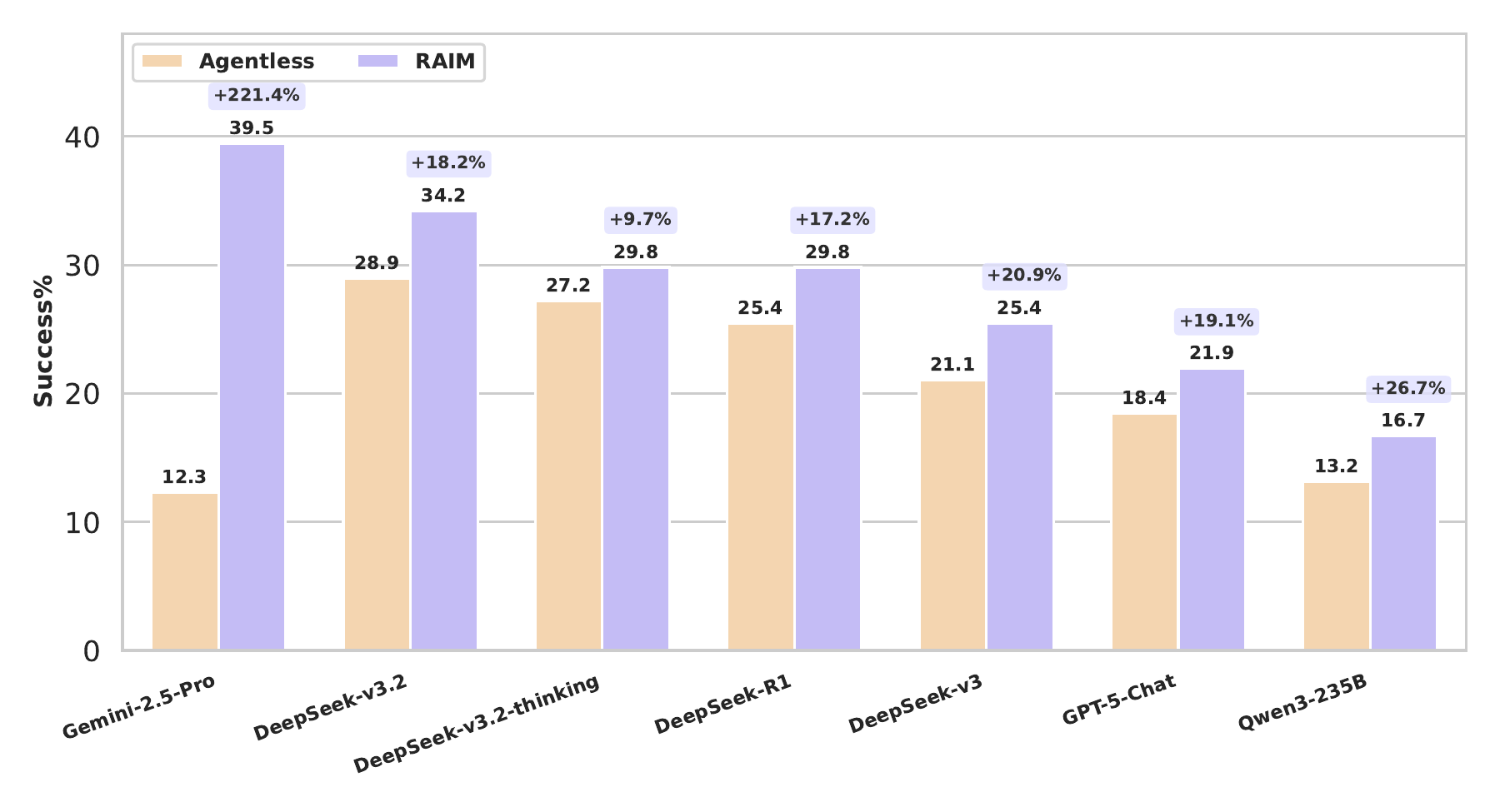}
  \caption{Performance comparison between \MyMethod{} and Agentless across LLMs on NoCode-bench Verified.}
  \label{fig:generalization}
\end{figure}
\textbf{Comparison with state-of-the-art baselines.}  Overall, \MyMethod{} establishes a new state-of-the-art benchmark with a Success(\%) of 39.47\% when utilizing Gemini-2.5-Pro. This result represents a substantial relative improvement of \textbf{36.34\%} over the previous best-performing method, Agentless (28.95\% with DeepSeek-v3.2), and a \textbf{55.15\%} improvement over OpenHands (25.44\% with Claude-4-Sonnet). This demonstrates that patch selection strategies based on multi-design and change impact analysis can enhance success rates for repository-level feature addition tasks. While larger language models generally exhibit greater capabilities, experimental results indicate that effective patch selection mechanisms remain crucial for feature addition tasks. Notably, \MyMethod{} achieved a 34.21\% success rate on the open-source model DeepSeek-v3.2, outperforming several competing methods utilizing more powerful base models. For instance, OpenHands and Agentless, both employing Claude-4-Sonnet, achieved 25.44\% and 28.07\% success rates respectively. These comparisons demonstrate that improvements in model architecture alone are insufficient to guarantee performance gains. Instead, function localization guided by code graphs and patch selection strategies based on multi-design and change impact analysis can compensate or even surpass the advantages derived from model scale. Furthermore, compared to Agentless's strategy of generating unplanned patches with 40 samples per instance, \MyMethod{} achieves higher success rates in feature addition by generating only 5 designs and patches.

\textbf{Comparison across different LLM backbones.} 
To evaluate the generalization capability of \MyMethod{} across diverse LLMs, performance is compared when the framework is instantiated with multiple representative LLMs. Agentless is selected as the primary point of comparison for this analysis because it demonstrates superior overall performance compared to OpenHands, thereby serving as a more competitive and representative baseline. As shown in Fig.~\ref{fig:generalization}, \MyMethod{} consistently outperforms Agentless across all evaluated models, which demonstrates its effectiveness independent of specific model or infrastructure choices.

Compared to the baseline method, \MyMethod{} achieves performance gains ranging from 9.7\% to 221.4\%. Notably, on the Gemini-2.5-Pro model, the baseline attains only a 12.3\% success rate, whereas \MyMethod{} substantially improves it to 39.5\%, indicating that our framework effectively unlocks the reasoning potential of the underlying LLM. Moreover, even on the strongest baseline model, DeepSeek-v3.2, \MyMethod{} still delivers a 18.2\% improvement, raising the success rate to 34.2\%. These results suggest that \MyMethod{} exhibits strong robustness and generalization across diverse LLM backbones.


\finding{\MyMethod{} establishes a state-of-the-art success rate of 39.47\% on NoCode-bench Verified, representing a \textbf{36.34\% relative improvement} over the best baseline, while enabling the open-source DeepSeek-v3.2 to outperform competitors utilizing powerful closed-source models. Furthermore, \MyMethod{} demonstrates robust generalization by consistently surpassing Agentless across seven diverse LLMs, yielding relative performance gains ranging from 9.7\% to 221.4\%.}

\subsection{RQ2: Cross-File Modification Performance Analysis}

\subsubsection{Design}
To evaluate the effectiveness of \MyMethod{} in handling cross-file modification tasks, the NoCode-bench Verified dataset is stratified based on the scope of modifications required by the ground-truth patches. Specifically, the 114 tasks are categorized into two distinct subsets: \textbf{single-file modifications} (61 instances) and \textbf{multi-file modifications} (53 instances). Subsequently, we computed and compared the success rates of \MyMethod{} against baseline frameworks Agentless and OpenHands across multiple LLMs. This segmentation enables a concrete evaluation of whether the proposed method mitigates the cross-file editing limitations prevalent in existing LLM-based repository-level feature addition frameworks.
\begin{table}[htbp]
\footnotesize
\renewcommand{\arraystretch}{0.8}
  \centering
  \caption{Success rate comparison across different file modification types on NoCode-bench Verified.}
  \label{tab:file_modification_rates}
  \begin{tabular}{lcccc}
    \toprule
    \textbf{Method} & \textbf{Model} & \textbf{Single File(\%)} & \textbf{Multi File(\%)} & \textbf{Overall(\%)} \\
    \midrule
    \multirow{5}{*}{OpenHands} 
        & Qwen3-235B      & 13.11 & 1.89 & 7.89 \\
        & DeepSeek-R1     & 11.48 & 1.89 & 7.02 \\
        & DeepSeek-v3     & 14.75 & 7.55 & 11.40 \\
        & Gemini-2.5-Pro  & 0.00 & 0.00 & 0.00 \\
        & Claude-4-Sonnet & 39.34 & 9.43 & 25.44 \\
    \midrule
    \multirow{8}{*}{Agentless} 
        & Qwen3-235B             & 21.31 & 3.77 & 13.16 \\
        & GPT-5-Chat             & 27.87 & 7.55 & 18.42 \\
        & DeepSeek-R1            & 40.98 & 7.55 & 25.44 \\
        & DeepSeek-v3            & 31.15 & 9.43 & 21.05 \\
        & DeepSeek-v3.2          & 47.54 & 7.55 & 28.95 \\
        & DeepSeek-v3.2-thinking & 39.34 & 13.21 & 27.19 \\
        & Gemini-2.5-Pro         & 19.67 & 3.77 & 12.28 \\
        & Claude-4-Sonnet        & 44.26 & 9.43 & 28.07 \\

    \midrule
    \multirow{7}{*}{\MyMethod{}} 
        & Qwen3-235B             & 26.23 & 5.66 & 16.67 \\
        & GPT-5-Chat             & 29.51 & 13.21 & 21.93 \\
        & DeepSeek-v3            & 39.34 & 9.43 & 25.44 \\
        & DeepSeek-R1            & 44.26 & 13.21 & 29.82 \\
        & DeepSeek-v3.2          & 49.18 & 16.98 & 34.21 \\
        & DeepSeek-v3.2-thinking & 44.26 & 13.21 & 29.82 \\
        & Gemini-2.5-Pro         & \textbf{57.38} & \textbf{18.87} & \textbf{39.47} \\

    \bottomrule
  \end{tabular}
\end{table}
\subsubsection{Results}
\textbf{Performance Comparison across Modification Scopes.}
Table \ref{tab:file_modification_rates} demonstrates the performance comparison of different methods in single-file and multi-file modification tasks. Results reveal that baseline methods exhibit a sharp performance decline when transitioning from single-file to multi-file editing, with success rates often plummeting to single digits. This disparity exposes the limitations of baseline methods in handling structural and logical dependencies across multiple files. Consequently, they struggle with tasks where code changes are interrelated yet dispersed across different modules, leading to incomplete or inconsistent implementation outcomes. In contrast, experimental results demonstrate that \MyMethod{} achieves significant improvements in processing capability within this critical domain.



\textbf{Enhanced Capability in Cross-File Modifications.}
In comparison to baseline methods, \MyMethod{} exhibits substantial performance gains when addressing multi-file tasks. As shown in Table \ref{tab:file_modification_rates}, Agentless typically struggles with multi-file editing, with success rates generally hovering between 3\% and 9\%. For instance, an agentless framework using Gemini-2.5-Pro achieved only a 3.77\% success rate in multi-file tasks, whereas \MyMethod{} utilizing the same model improved this figure to 18.87\%, representing a relative increase of 400.5\%. Similarly, when using DeepSeek-v3.2, \MyMethod{} achieved a success rate of 16.98\%, while the agentless framework managed only 7.55\%, representing a relative improvement of 124.9\%. This demonstrates that the code graph search strategy within \MyMethod{} can precisely locate dispersed code regions. The collaboration between multi-design generation and change impact analysis expands the solution space while validating the structural consistency of cross-file patches. This ensures selected patches correctly handle complex structural dependencies, effectively mitigating the inherent preference of LLMs for localized single-file edits.

\textbf{Robustness in Single-File Tasks.}
Beyond excelling in complex multi-file scenarios, \MyMethod{} maintains or even enhances performance in single-file tasks. When employing the Gemini-2.5-Pro model, this method achieves a 57.38\% success rate in single-file modification tasks. Compared to the Agentless framework using the same model, which achieved only 19.67\%, this represents a significant relative improvement of 191.7\%. Furthermore, \MyMethod{} substantially outperforms the best results of all baseline methods, such as the 47.54\% achieved by Agentless with DeepSeek-v3.2, establishing a relative lead of 20.7\%. This demonstrates that the multi-design patch generation and code change impact analysis strategy does not hinder performance on simple tasks. Instead, these mechanisms provide effective validation, ensuring local changes strictly align with the project's overall context and thereby effectively reducing regression errors.

\finding{\MyMethod{} mitigates cross-file editing bottlenecks by achieving an 18.87\% success rate in multi-file tasks, marking a 191.7\% relative improvement over Agentless. This underscores its superior ability to coordinate complex and interdependent modifications across the repository.}

\subsection{RQ3: Ablation}
\subsubsection{Design}
To isolate the influence of key components on the success rate of feature addition, an ablation study is conducted using DeepSeek-v3.2 as the underlying model. Three variants are evaluated, each modifying a single module while preserving the rest of the pipeline:
\begin{itemize}
    \item \rev{\textbf{w/o Code Graph}: the multi-round graph-based search is replaced by directly providing the LLM with the feature description and the code skeletons of identified feature-relevant files for function-level localization.}
    \item \rev{\textbf{w/o Multi-Design}: without the multi-design generation, a single patch is generated directly.}
    \item \rev{\textbf{w/o Patch Evaluation}: directly using LLM for subsequent patch selection without patch evaluation based on change impact analysis.}
\end{itemize}


\begin{table}[htbp]
\footnotesize
\renewcommand{\arraystretch}{0.8}
\centering
\caption{Ablation study of \MyMethod{} on NoCode-bench Verified.}
\label{tab:ablation_study}
\begin{tabular}{lcccc}
\toprule
\textbf{Method} & \textbf{Success(\%)} & \textbf{RT(\%)} & \textbf{FV-Micro(\%)} & \textbf{FV-Macro(\%)} \\
\midrule

w/o Code Graph 
& 29.82\,\drop{12.83} 
& 85.09\,\drop{1.01} 
& 15.99\,\drop{0.12} 
& 41.51\,\drop{8.93} \\

w/o Multi-Design 
& 21.05\,\drop{38.47} 
& 70.18\,\drop{18.36} 
& 5.95\,\drop{62.84} 
& 30.18\,\drop{33.79} \\

w/o Patch Evaluation 
& 22.81\,\drop{33.32} 
& 77.19\,\drop{10.20} 
& 7.53\,\drop{52.97} 
& 33.68\,\drop{26.11} \\

\midrule
\textbf{\MyMethod{}} 
& \textbf{34.21} 
& \textbf{85.96} 
& \textbf{16.01} 
& \textbf{45.58} \\
\bottomrule
\end{tabular}
\end{table}

\subsubsection{Results}
Table~\ref{tab:ablation_study} demonstrates the performance of each \MyMethod{} variant relative to the \textbf{complete framework}. The results consistently show that removing any key component leads to comprehensive declines in success rate and feature validation metrics, which confirms the synergistic contribution of these modules to the overall effectiveness. Notably, the removal of the \textbf{multi-design generation strategy} results in the most substantial performance degradation, specifically a 38.47\% drop in success rate and a 62.84\% decline in micro-level functional validation metrics. This discrepancy indicates that without a \textbf{structured multi-design process}, the model may locate relevant code regions but fails to synthesize logically correct feature implementations, which identifies the \textbf{multi-design strategy} as a core critical factor. Similarly, excluding the patch evaluation component causes a substantial 33.32\% drop in success rate, which demonstrates that assessment based on change impact analysis is essential for filtering out seemingly plausible yet erroneous solutions. Finally, the removal of the code graph-based function localization leads to a 12.83\% decrease in success rate. This finding suggests that while the \textbf{multi-design} and evaluation phases drive high-level reasoning, the structural localization provided by the code graph remains a critical foundation that ensures subsequent phases operate within an accurate and relevant context.

\finding{Multi-design generation and patch evaluation serve as the primary drivers of performance, with their removal causing substantial success rate declines of 38.47\% and 33.32\%, respectively. Furthermore, the omission of the code graph leads to a 12.83\% drop in effectiveness, which confirms its essential role in ensuring accurate localization and context retrieval.}

\subsection{RQ4: Effectiveness of Multi-Design and Selection Strategies}

\subsubsection{Design}
To evaluate the effectiveness of the multi-design-based patch generation and code change impact analysis-based patch selection strategies, the results of patch selection for \MyMethod{} across seven distinct LLMs are statistically analyzed. Four evaluation metrics are utilized: the solvable instance count (SC), which represents the number of tasks where at least one design yields a correct implementation; the match count (MC), reflecting the instances where the selector correctly identifies a successful patch; the patch selection accuracy (Acc(\%)); and the final success rate (Success(\%)). Additionally, a parameter sensitivity analysis is performed on DeepSeek-v3.2 to investigate the influence of design quantity, as this focused approach provides clear insights while managing the high computational cost of the experiment.

\begin{table}[htbp]
\footnotesize
\renewcommand{\arraystretch}{0.8}
\centering
\caption{Performance of multi-design patch generation and patch selection strategies across different LLMs.}
\label{tab:multi_model_analysis}
\begin{tabular}{lcccc}
\toprule
\textbf{Model} & \textbf{SC} & \textbf{MC} & \textbf{Acc(\%)} & \textbf{Success(\%)} \\
\midrule
Qwen3-235B & 27 & 19 & 70.37 & 16.67 \\
GPT-5-Chat & 28 & 23 & 82.14 & 20.18 \\
DeepSeek-v3 & 38 & 29 & 76.32 & 25.44 \\
DeepSeek-R1 & 39 & 34 & 87.18 & 29.82 \\
DeepSeek-v3.2 & 44 & 39 & \textbf{88.64} & 34.21 \\
DeepSeek-v3.2-thinking & 44 & 34 & 77.27 & 29.82 \\
Gemini-2.5-Pro & \textbf{54} & \textbf{45} & 83.33 & \textbf{39.47} \\
\bottomrule
\end{tabular}
\end{table}

\begin{table}[htbp]
\footnotesize
\renewcommand{\arraystretch}{0.8}
\centering
\caption{Impact of design quantity on selection accuracy and overall success.}
\label{tab:multi_plan_analysis}
\begin{tabular}{lccccc}
\toprule
\textbf{Model} & \textbf{Design Count} & \textbf{SC} & \textbf{MC} & \textbf{Acc(\%)} & \textbf{Success(\%)} \\
\midrule
\multirow{4}{*}{DeepSeek-v3.2} & 3 & 42 & 36 & 85.71 & 31.58 \\
                               & 5 & 44 & \textbf{39} & \textbf{88.64} & \textbf{34.21} \\
                               & 7 & \textbf{48} & 37 & 77.08 & 32.46 \\
                               & 9 & 46 & 33 & 71.74 & 28.95 \\
\bottomrule
\end{tabular}
\end{table}
\subsubsection{Results}

\textbf{Multi-Design Performance Analysis.}
Table~\ref{tab:multi_model_analysis} presents the performance of the multi-design-based patch generation and code change impact analysis-based patch selection strategies. The results demonstrate the effectiveness of \MyMethod{} from two critical perspectives. First, the high SC values indicate that the multi-design-based patch generation strategy effectively diversifies the solution space. For example, Gemini-2.5-Pro achieves a peak of 54 solvable instances and DeepSeek-v3.2 reaches 44. This capability significantly raises the upper bound of potential success by exploring a wider solution space. Second, the high selection accuracy confirms that the selection mechanism based on change impact analysis accurately identifies the optimal patch. With accuracy consistently exceeding 70\% across all models and reaching up to 88.64\%, the framework successfully converts the potential offered by multi-design planning into actual performance.

\textbf{Parameter Sensitivity Analysis.}
Table~\ref{tab:multi_plan_analysis} further examines the impact of varying the number of designs on performance using DeepSeek-v3.2. It reveals a clear trade-off between task coverage and selection precision. When the design count increases from 3 to 5, the success rate peaks at 34.21\% and Acc\% reaches its highest point at 88.64\%. Although increasing designs to 7 yields a higher SC of 48, the selection accuracy declines significantly. This drop indicates that an excessive number of candidates introduces interference noise, which complicates the evaluation process and diminishes selection reliability. Therefore, a configuration of 5 designs achieves the optimal balance by maintaining high precision while ensuring sufficient solution diversity.

\finding{Synergizing multi-design-based patch generation with code change impact analysis-based patch selection effectively translates expanded implementation diversity into actual success, achieving over 70\% accuracy across leading models. Furthermore, generating five designs provides an optimal balance between solution coverage and selection precision.}

\subsection{RQ5: Effectiveness in Mitigating Common Baseline Failures}
\subsubsection{Design}
To evaluate how \MyMethod{} mitigates typical shortcomings, failure mode distributions are analyzed alongside a successful implementation. Failures are categorized into \textbf{Regression Errors} (P2P test failures), denoting disruption of existing functionality, and \textbf{New Feature Errors} (F2P test failures), signifying incorrect requirements implementation. This classification assesses the balance between feature completion and system stability, while a case study on \texttt{pylint-7869} illustrates the practical advantages of architectural awareness over localized editing.

\begin{table}[htbp]
\centering
\footnotesize
\renewcommand{\arraystretch}{0.8}
\caption{Comparison of failure type distributions between \MyMethod{} and baseline methods across different LLMs.}
\label{tab:failure_analysis_optimized}
\begin{tabular}{llccc}
\toprule
\textbf{Model} & \textbf{Method} & \textbf{Success(\%)} & \textbf{Regression Error(\%)} & \textbf{New Feature Error(\%)} \\
\midrule
\multirow{3}{*}{DeepSeek-R1} 
 & OpenHands & 7.02 & 53.51 \rise{134.59} & 92.11 \rise{36.38} \\
 & Agentless & 25.44 & 20.18 \drop{11.53} & 71.93 \rise{6.50} \\
 & \textbf{\MyMethod} & 29.82 & 22.81 & 67.54 \\
\midrule
\multirow{3}{*}{DeepSeek-v3} 
 & OpenHands & 11.40 & 50.88 \rise{176.22} & 85.96 \rise{18.06} \\
 & Agentless & 21.05 & 19.30 \rise{4.78} & 76.32 \rise{4.82} \\
 & \textbf{\MyMethod} & 25.44 & 18.42 & 72.81 \\
\midrule
\multirow{3}{*}{Gemini-2.5-Pro} 
 & OpenHands & 0.00 & 38.60 \rise{120.07} & 100.00 \rise{72.74} \\
 & Agentless & 12.28 & 25.44 \rise{45.04} & 86.84 \rise{50.01} \\
 & \textbf{\MyMethod} & 39.47 & 17.54 & 57.89 \\
\midrule
\multirow{3}{*}{Qwen3-235B} 
 & OpenHands & 7.89 & 52.63 \rise{172.69} & 91.23 \rise{11.83} \\
 & Agentless & 13.16 & 23.68 \rise{22.69} & 85.09 \rise{4.30} \\
 & \textbf{\MyMethod} & 16.67 & 19.30 & 81.58 \\
\bottomrule
\end{tabular}
\end{table}

\begin{figure}[htbp]
  \centering
  \includegraphics[width=0.9\linewidth]{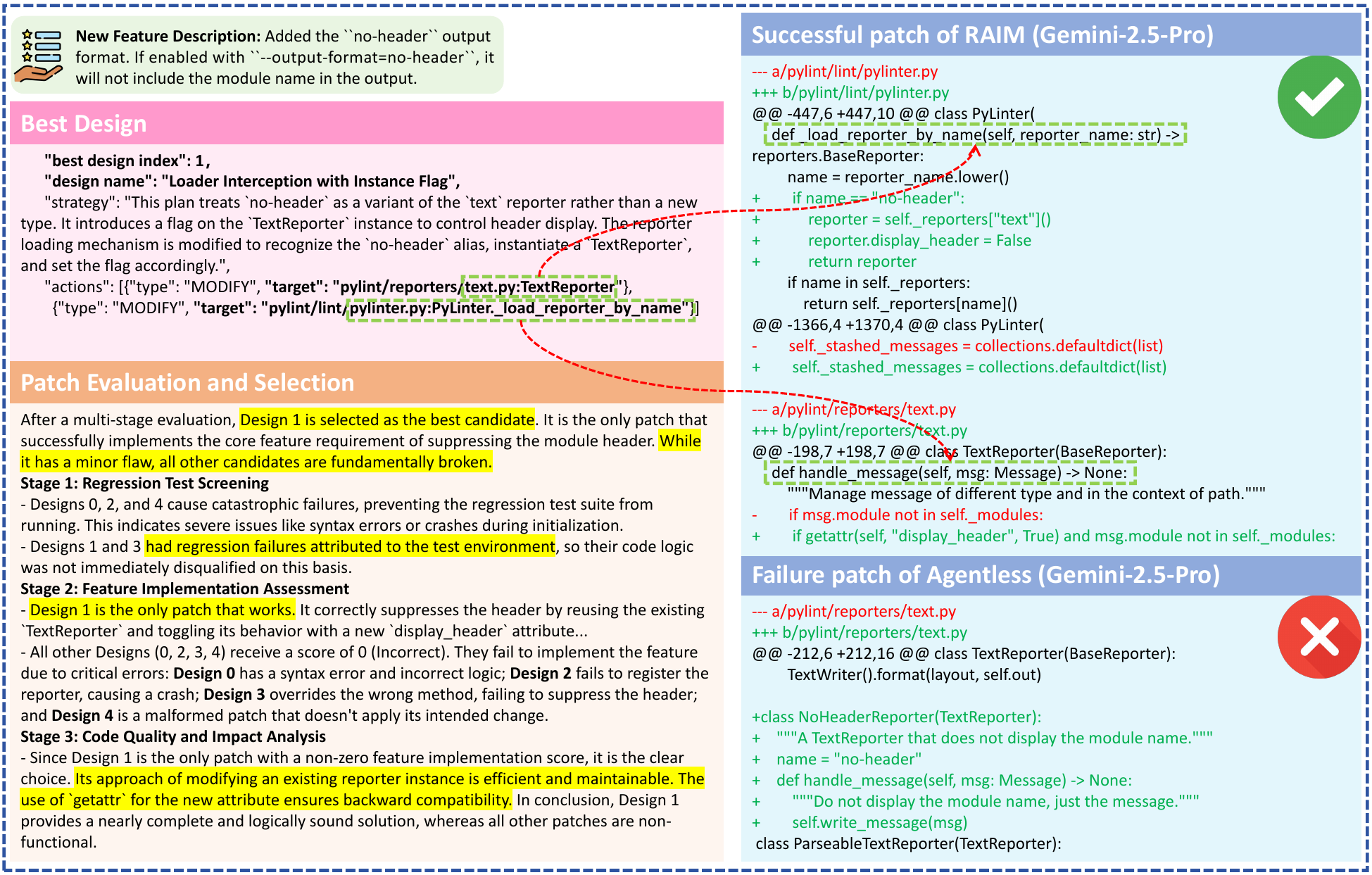}
  \caption{Case study of \MyMethod{} with instance \textit{pylint-dev\_\_pylint-7869}.}
  \label{fig:case_study}
\end{figure}
\subsubsection{Results}
\textbf{Quantitative analysis of failure types.}
Table~\ref{tab:failure_analysis_optimized} illustrates that \MyMethod{} consistently delivers superior performance by effectively mitigating both regression and implementation errors across various architectures. The significant reduction in New Feature Errors across all models is attributed to the code graph-based multi-round localization and call-graph contextual enrichment, which provide the LLM with sufficient architectural awareness to identify the precise modification points. Simultaneously, \MyMethod{} demonstrates a strong capability in minimizing Regression Errors. This improvement stems from the multi-design-based patch generation and patch selection mechanism, which rigorously evaluates candidate patches through static dependency tracing and dynamic testing to identify and discard implementations that disrupt legacy logic. For instance, when utilizing the Gemini-2.5-Pro model, \MyMethod{} reduces regression errors by 45.04\% and new feature errors by 50.01\% compared to Agentless. These results confirm that the framework successfully balances functional progress with system stability by selecting the most architecturally consistent implementation from a diverse pool of designs.

A minor exception occurs with DeepSeek-R1, where the model occasionally produces structurally complex designs that introduce subtle regressions not present in the more conservative baseline. However, the significantly lower New Feature Error rate on this model still underscores the superior capacity of \MyMethod{} in achieving functional completeness for complex tasks.

\textbf{Qualitative case study.}
The effectiveness of \MyMethod{} in resolving complex dependencies is further validated through the \texttt{pylint-7869} instance, which requires implementing a \textit{no-header} output format. As shown in Fig.~\ref{fig:case_study}, Agentless fails because it produces an isolated implementation. Although the baseline correctly defines the \texttt{NoHeaderReporter} class, it overlooks the architectural necessity of registering the new component within the system loader in \texttt{pylinter.py}.

In contrast, \MyMethod{} successfully identifies the optimal implementation pathway through its multi-design generation process. By leveraging the code graph to explore diverse strategies, the framework selects the \textit{Loader Interception} design. This design introduces a defensive flag within the existing reporter and updates the loading logic to toggle its state. This approach ensures the new feature is fully integrated into the system control flow while strictly maintaining the original behavior for existing code paths, thereby achieving functional correctness without architectural disruption.

\finding{\MyMethod{} addresses common failure modes by ensuring that new features are integrated into the existing system architecture rather than implemented in isolation. By combining multi-design generation with rigorous impact assessment, the framework effectively minimizes both regression risks and functional implementation errors in complex repository-level tasks.}

\section{Related Work}
\label{sec:realted work}
\subsection{Feature Localization and Code Recommendation}
Feature localization and recommendation facilitate software evolution \cite{niu2025feature,liu2024large,tao2025retrieval,yang2025survey}. Early systems match requests to APIs via historical changes \cite{thung2013automatic}, similarity integration \cite{xu2018mulapi}, or CNN-mapping \cite{sun2019enabling}. Beyond suggestions, automated methods support structural preparation through refactoring \cite{nyamawe2019automated}, traceability mining \cite{licorish2018linking,palomba2015user}, and opportunity prediction \cite{nyamawe2022identifying,nyamawe2020feature}.

Feature localization has evolved from traditional IR-based~\cite{marcus2004information,zhou2012should,xue2012feature} and Spectrum-based~\cite{dusia2016recent,abreu2007accuracy} techniques to modern semantic approaches. Recent works leverage hybrid methods combining user feedback~\cite{palomba2017recommending}, word embeddings~\cite{zhang2019where2change}, and structural connections~\cite{chiyourcolo} to improve accuracy. While methods like URR~\cite{ciurumelea2017analyzing} and RISING~\cite{zhou2020user} categorize reviews for better targeting, they primarily focus on recommendation rather than the autonomous generation required by NoCode-bench~\cite{deng2025nocode}.

\subsection{Repository-Level Issue Resolution}

Benchmarks like SWE-bench \cite{jimenez2023swe} have shifted research focus toward repository-level maintenance, involving autonomous agents and structured pipelines \cite{arora2024masai,ouyang2024repograph,chen2025locagent,phan2024hyperagent,tao2024magis}. Agent-based frameworks, pioneered by SWE-agent \cite{yang2024swe} and RepairAgent \cite{bouzenia2024repairagent}, have evolved into versatile platforms such as OpenHands \cite{wang2024openhands}, TRAE \cite{gao2025trae}, and ExecutionAgent \cite{bouzenia2025you} that prioritize collaborative intelligence and execution reliability. Conversely, workflow-based methods like Agentless \cite{xia2024agentless}, CodeMonkeys \cite{ehrlich2025codemonkeys}, and PatchPilot \cite{li2025patchpilot} provide streamlined localization-and-repair by optimizing specific sub-tasks without complex agentic loops. To capture interdependent dependencies, graph-enhanced methods such as RepoGraph \cite{ouyang2024repograph}, KGCompass \cite{yang2025enhancing}, and CodexGraph \cite{liu2025codexgraph} utilize repository-level structural knowledge. Further refinements include graph-based localization via LocAgent \cite{chen2025locagent} and OrcaLoca \cite{yu2025orcaloca}, alongside strategic solution exploration via SWE-Search \cite{antoniades2024swe} based on Monte Carlo Tree Search.

Despite these advancements, agents still struggle with complex modification patterns \cite{chen2025unveiling,zhang2024diversity}. A gap exists between defect resolution and feature addition formalized by NoCode-bench \cite{deng2025nocode}. Most methods are optimized for passing failing tests rather than designing architecturally consistent implementations. They lack systematic change impact analysis to evaluate side effects on legacy functionality \cite{deng2025nocode}. \MyMethod{} addresses this by integrating multi-design generation with an impact-aware patch selection, ensuring robust and regression-free features.

\section{Threats To Validity}
\label{sec:threats}
Several potential threats to the validity of our findings are considered and mitigated. \textbf{Internal validity} may be concerned with data contamination, as the open-source repositories in NoCode-bench might have been included in the pre-training corpora of the evaluated LLMs. We address this by emphasizing that \MyMethod{} relies on repository-level architectural reasoning and multi-design selection rather than simple code recall; the consistent performance gains across seven diverse models suggest that the improvements stem from our framework's logic rather than memorization. Regarding model-specific bias, we evaluate \MyMethod{} on both open-source and closed-source models to ensure the generalizability of our strategies. For \textbf{external validity}, although our evaluation focuses on Python projects within the NoCode-bench Verified dataset, the underlying principles of code-graph-based localization and change impact analysis are fundamentally language-agnostic and designed to scale to other complex software ecosystems. Finally, to ensure construct validity, we utilize standard metrics and follow the evaluation protocols established by prior high-impact studies, ensuring that our results are both rigorous and comparable.

\section{Conclusion and Future Work}
\label{sec:conclusion}
In this paper, \MyMethod{} is proposed as a repository-level architecture-aware framework for natural language-driven feature addition. By synergizing architecture-aware feature localization with impact-aware multi-design patch generation, the framework overcomes the architectural blindness and fragility of current workflow-based and agent-based approaches. \MyMethod{} establishes a new state-of-the-art success rate of 39.47\% on NoCode-bench Verified, marking a 36.34\% relative improvement over the best-performing baseline. These results highlight the necessity of structural awareness and impact assessment for maintaining architectural consistency during software evolution. Future research will explore multi-step feature evolution and autonomous self-repair mechanisms to further enhance the reliability and precision of implementation designs in complex environments.

\section{DATA AVAILABILITY}
To facilitate the replication study, we have released our data and code at:\url{https://github.com/Ccsosad/RAIM}. 


\bibliographystyle{unsrt}  
\bibliography{references}

@String{Computer = "{IEEE} Computer" }

@String{Academic = "Academic Press" }

@String{Springer = "Springer-Verlag" }

@article{deng2025nocode,
  title={Nocode-bench: A benchmark for evaluating natural language-driven feature addition},
  author={Deng, Le and Jiang, Zhonghao and Cao, Jialun and Pradel, Michael and Liu, Zhongxin},
  journal={arXiv preprint arXiv:2507.18130},
  year={2025}
}

@article{xia2024agentless,
  title={Agentless: Demystifying llm-based software engineering agents},
  author={Xia, Chunqiu Steven and Deng, Yinlin and Dunn, Soren and Zhang, Lingming},
  journal={arXiv preprint arXiv:2407.01489},
  year={2024}
}

@article{wang2024openhands,
  title={Openhands: An open platform for ai software developers as generalist agents},
  author={Wang, Xingyao and Li, Boxuan and Song, Yufan and Xu, Frank F and Tang, Xiangru and Zhuge, Mingchen and Pan, Jiayi and Song, Yueqi and Li, Bowen and Singh, Jaskirat and others},
  journal={arXiv preprint arXiv:2407.16741},
  year={2024}
}

@article{chen2021evaluating,
  title={Evaluating large language models trained on code},
  author={Chen, Mark},
  journal={arXiv preprint arXiv:2107.03374},
  year={2021}
}

@article{jimenez2023swe,
  title={Swe-bench: Can language models resolve real-world github issues?},
  author={Jimenez, Carlos E and Yang, John and Wettig, Alexander and Yao, Shunyu and Pei, Kexin and Press, Ofir and Narasimhan, Karthik},
  journal={arXiv preprint arXiv:2310.06770},
  year={2023}
}

@article{hirzel2023low,
  title={Low-code programming models},
  author={Hirzel, Martin},
  journal={Communications of the ACM},
  volume={66},
  number={10},
  pages={76--85},
  year={2023},
  publisher={ACM New York, NY, USA}
}

@inproceedings{rao2024ai,
  title={AI for Low-Code for AI},
  author={Rao, Nikitha and Tsay, Jason and Kate, Kiran and Hellendoorn, Vincent and Hirzel, Martin},
  booktitle={Proceedings of the 29th International Conference on Intelligent User Interfaces},
  pages={837--852},
  year={2024}
}

@article{yin2017syntactic,
  title={A syntactic neural model for general-purpose code generation},
  author={Yin, Pengcheng and Neubig, Graham},
  journal={arXiv preprint arXiv:1704.01696},
  year={2017}
}

@article{glass2001frequently,
  title={Frequently forgotten fundamental facts about software engineering},
  author={Glass, Robert L},
  journal={IEEE software},
  volume={18},
  number={3},
  pages={112--111},
  year={2001}
}

@inproceedings{rahman2019supporting,
  title={Supporting code search with context-aware, analytics-driven, effective query reformulation},
  author={Rahman, Mohammad Masudur},
  booktitle={2019 IEEE/ACM 41st International Conference on Software Engineering: Companion Proceedings (ICSE-Companion)},
  pages={226--229},
  year={2019},
  organization={IEEE}
}

@article{ouyang2024repograph,
  title={Repograph: Enhancing ai software engineering with repository-level code graph},
  author={Ouyang, Siru and Yu, Wenhao and Ma, Kaixin and Xiao, Zilin and Zhang, Zhihan and Jia, Mengzhao and Han, Jiawei and Zhang, Hongming and Yu, Dong},
  journal={arXiv preprint arXiv:2410.14684},
  year={2024}
}

@article{ruan2024specrover,
  title={Specrover: Code intent extraction via llms},
  author={Ruan, Haifeng and Zhang, Yuntong and Roychoudhury, Abhik},
  journal={arXiv preprint arXiv:2408.02232},
  year={2024}
}

@article{niu2025feature,
  title={Feature Request Analysis and Processing: Tasks, Techniques, and Trends},
  author={Niu, Feifei and Li, Chuanyi and Zuo, Haosheng and Wu, Jionghan and Xia, Xin},
  journal={arXiv preprint arXiv:2508.12436},
  year={2025}
}

@inproceedings{thung2013automatic,
  title={Automatic recommendation of API methods from feature requests},
  author={Thung, Ferdian and Wang, Shaowei and Lo, David and Lawall, Julia},
  booktitle={2013 28th IEEE/ACM International Conference on Automated Software Engineering (ASE)},
  pages={290--300},
  year={2013},
  organization={IEEE}
}

@article{xu2018mulapi,
  title={MULAPI: Improving API method recommendation with API usage location},
  author={Xu, Congying and Sun, Xiaobing and Li, Bin and Lu, Xintong and Guo, Hongjing},
  journal={Journal of Systems and Software},
  volume={142},
  pages={195--205},
  year={2018},
  publisher={Elsevier}
}

@article{sun2019enabling,
  title={Enabling feature location for API method recommendation and usage location},
  author={Sun, Xiaobing and Xu, Congying and Li, Bin and Duan, Yucong and Lu, Xintong},
  journal={IEEE Access},
  volume={7},
  pages={49872--49881},
  year={2019},
  publisher={IEEE}
}

@inproceedings{nyamawe2019automated,
  title={Automated recommendation of software refactorings based on feature requests},
  author={Nyamawe, Ally S and Liu, Hui and Niu, Nan and Umer, Qasim and Niu, Zhendong},
  booktitle={2019 IEEE 27th International Requirements Engineering Conference (RE)},
  pages={187--198},
  year={2019},
  organization={IEEE}
}

@inproceedings{marcus2004information,
  title={An information retrieval approach to concept location in source code},
  author={Marcus, Andrian and Sergeyev, Andrey and Rajlich, Vaclav and Maletic, Jonathan I},
  booktitle={11th working conference on reverse engineering},
  pages={214--223},
  year={2004},
  organization={IEEE}
}

@inproceedings{palomba2017recommending,
  title={Recommending and localizing change requests for mobile apps based on user reviews},
  author={Palomba, Fabio and Salza, Pasquale and Ciurumelea, Adelina and Panichella, Sebastiano and Gall, Harald and Ferrucci, Filomena and De Lucia, Andrea},
  booktitle={2017 IEEE/ACM 39th International Conference on Software Engineering (ICSE)},
  pages={106--117},
  year={2017},
  organization={IEEE}
}

@article{zhang2019where2change,
  title={Where2Change: Change request localization for app reviews},
  author={Zhang, Tao and Chen, Jiachi and Zhan, Xian and Luo, Xiapu and Lo, David and Jiang, He},
  journal={IEEE Transactions on Software Engineering},
  volume={47},
  number={11},
  pages={2590--2616},
  year={2019},
  publisher={IEEE}
}

@article{chiyourcolo,
  title={YourCoLo: Leveraging One-to-Many Relationships and Inter-Code Connections for User Review-Based Code Localization},
  author={Chi, Kuo and Niu, Changan and Yang, Zhou and Li, Chuanyi and Feng, Yi and Ge, Jidong and Luo, Bin and Lo, David and Ng, Vincent},
  journal={ACM Transactions on Software Engineering and Methodology},
  publisher={ACM New York, NY}
}

@inproceedings{ciurumelea2017analyzing,
  title={Analyzing reviews and code of mobile apps for better release planning},
  author={Ciurumelea, Adelina and Schaufelb{\"u}hl, Andreas and Panichella, Sebastiano and Gall, Harald C},
  booktitle={2017 IEEE 24th international conference on software analysis, evolution and reengineering (SANER)},
  pages={91--102},
  year={2017},
  organization={IEEE}
}

@article{yang2024swe,
  title={Swe-agent: Agent-computer interfaces enable automated software engineering},
  author={Yang, John and Jimenez, Carlos E and Wettig, Alexander and Lieret, Kilian and Yao, Shunyu and Narasimhan, Karthik and Press, Ofir},
  journal={Advances in Neural Information Processing Systems},
  volume={37},
  pages={50528--50652},
  year={2024}
}

@article{bouzenia2024repairagent,
  title={Repairagent: An autonomous, llm-based agent for program repair},
  author={Bouzenia, Islem and Devanbu, Premkumar and Pradel, Michael},
  journal={arXiv preprint arXiv:2403.17134},
  year={2024}
}

@article{gao2025trae,
  title={Trae agent: An llm-based agent for software engineering with test-time scaling},
  author={Gao, Pengfei and Tian, Zhao and Meng, Xiangxin and Wang, Xinchen and Hu, Ruida and Xiao, Yuanan and Liu, Yizhou and Zhang, Zhao and Chen, Junjie and Gao, Cuiyun and others},
  journal={arXiv preprint arXiv:2507.23370},
  year={2025}
}

@article{bouzenia2025you,
  title={You name it, I run it: An LLM agent to execute tests of arbitrary projects},
  author={Bouzenia, Islem and Pradel, Michael},
  journal={Proceedings of the ACM on Software Engineering},
  volume={2},
  number={ISSTA},
  pages={1054--1076},
  year={2025},
  publisher={ACM New York, NY, USA}
}

@article{li2025patchpilot,
  title={PatchPilot: A Cost-Efficient Software Engineering Agent with Early Attempts on Formal Verification},
  author={Li, Hongwei and Tang, Yuheng and Wang, Shiqi and Guo, Wenbo},
  journal={arXiv preprint arXiv:2502.02747},
  year={2025}
}

@article{liu2024large,
  title={Large language model-based agents for software engineering: A survey},
  author={Liu, Junwei and Wang, Kaixin and Chen, Yixuan and Peng, Xin and Chen, Zhenpeng and Zhang, Lingming and Lou, Yiling},
  journal={arXiv preprint arXiv:2409.02977},
  year={2024}
}

@article{tao2025retrieval,
  title={Retrieval-Augmented Code Generation: A Survey with Focus on Repository-Level Approaches},
  author={Tao, Yicheng and Qin, Yao and Liu, Yepang},
  journal={arXiv preprint arXiv:2510.04905},
  year={2025}
}

@article{yang2025survey,
  title={A Survey of LLM-based Automated Program Repair: Taxonomies, Design Paradigms, and Applications},
  author={Yang, Boyang and Cai, Zijian and Liu, Fengling and Le, Bach and Zhang, Lingming and Bissyand{\'e}, Tegawend{\'e} F and Liu, Yang and Tian, Haoye},
  journal={arXiv preprint arXiv:2506.23749},
  year={2025}
}

@inproceedings{licorish2018linking,
  title={Linking User Requests, Developer Responses and Code Changes: Android OS Case Study},
  author={Licorish, Sherlock A and Zolduoarrati, Elijah and Stanger, Nigel},
  booktitle={Proceedings of the 22nd International Conference on Evaluation and Assessment in Software Engineering 2018},
  pages={79--89},
  year={2018}
}

@inproceedings{palomba2015user,
  title={User reviews matter! tracking crowdsourced reviews to support evolution of successful apps},
  author={Palomba, Fabio and Linares-V{\'a}squez, Mario and Bavota, Gabriele and Oliveto, Rocco and Di Penta, Massimiliano and Poshyvanyk, Denys and De Lucia, Andrea},
  booktitle={2015 IEEE international conference on software maintenance and evolution (ICSME)},
  pages={291--300},
  year={2015},
  organization={IEEE}
}

@article{nyamawe2022identifying,
  title={Identifying rename refactoring opportunities based on feature requests},
  author={Nyamawe, Ally S and Bakhti, Khadidja and Sandiwarno, Sulis},
  journal={International Journal of Computers and Applications},
  volume={44},
  number={8},
  pages={770--778},
  year={2022},
  publisher={Taylor \& Francis}
}

@article{nyamawe2020feature,
  title={Feature requests-based recommendation of software refactorings},
  author={Nyamawe, Ally S and Liu, Hui and Niu, Nan and Umer, Qasim and Niu, Zhendong},
  journal={Empirical Software Engineering},
  volume={25},
  number={5},
  pages={4315--4347},
  year={2020},
  publisher={Springer}
}

@inproceedings{zhou2012should,
  title={Where should the bugs be fixed? more accurate information retrieval-based bug localization based on bug reports},
  author={Zhou, Jian and Zhang, Hongyu and Lo, David},
  booktitle={2012 34th International conference on software engineering (ICSE)},
  pages={14--24},
  year={2012},
  organization={IEEE}
}

@inproceedings{xue2012feature,
  title={Feature location in a collection of product variants},
  author={Xue, Yinxing and Xing, Zhenchang and Jarzabek, Stan},
  booktitle={2012 19th Working Conference on Reverse Engineering},
  pages={145--154},
  year={2012},
  organization={IEEE}
}

@article{dusia2016recent,
  title={Recent advances in fault localization in computer networks},
  author={Dusia, Ayush and Sethi, Adarshpal S},
  journal={IEEE Communications Surveys \& Tutorials},
  volume={18},
  number={4},
  pages={3030--3051},
  year={2016},
  publisher={IEEE}
}

@inproceedings{abreu2007accuracy,
  title={On the accuracy of spectrum-based fault localization},
  author={Abreu, Rui and Zoeteweij, Peter and Van Gemund, Arjan JC},
  booktitle={Testing: Academic and industrial conference practice and research techniques-MUTATION (TAICPART-MUTATION 2007)},
  pages={89--98},
  year={2007},
  organization={IEEE}
}

@article{zhou2020user,
  title={User review-based change file localization for mobile applications},
  author={Zhou, Yu and Su, Yanqi and Chen, Taolue and Huang, Zhiqiu and Gall, Harald and Panichella, Sebastiano},
  journal={IEEE Transactions on Software Engineering},
  volume={47},
  number={12},
  pages={2755--2770},
  year={2020},
  publisher={IEEE}
}

@article{arora2024masai,
  title={Masai: Modular architecture for software-engineering ai agents},
  author={Arora, Daman and Sonwane, Atharv and Wadhwa, Nalin and Mehrotra, Abhav and Utpala, Saiteja and Bairi, Ramakrishna and Kanade, Aditya and Natarajan, Nagarajan},
  journal={arXiv preprint arXiv:2406.11638},
  year={2024}
}

@inproceedings{chen2025locagent,
  title={Locagent: Graph-guided llm agents for code localization},
  author={Chen, Zhaoling and Tang, Robert and Deng, Gangda and Wu, Fang and Wu, Jialong and Jiang, Zhiwei and Prasanna, Viktor and Cohan, Arman and Wang, Xingyao},
  booktitle={Proceedings of the 63rd Annual Meeting of the Association for Computational Linguistics (Volume 1: Long Papers)},
  pages={8697--8727},
  year={2025}
}

@article{phan2024hyperagent,
  title={Hyperagent: Generalist software engineering agents to solve coding tasks at scale},
  author={Phan, Huy Nhat and Nguyen, Tien N and Nguyen, Phong X and Bui, Nghi DQ},
  journal={arXiv preprint arXiv:2409.16299},
  year={2024}
}

@article{tao2024magis,
  title={Magis: Llm-based multi-agent framework for github issue resolution},
  author={Tao, Wei and Zhou, Yucheng and Wang, Yanlin and Zhang, Wenqiang and Zhang, Hongyu and Cheng, Yu},
  journal={Advances in Neural Information Processing Systems},
  volume={37},
  pages={51963--51993},
  year={2024}
}

@article{ehrlich2025codemonkeys,
  title={Codemonkeys: Scaling test-time compute for software engineering},
  author={Ehrlich, Ryan and Brown, Bradley and Juravsky, Jordan and Clark, Ronald and R{\'e}, Christopher and Mirhoseini, Azalia},
  journal={arXiv preprint arXiv:2501.14723},
  year={2025}
}

@inproceedings{liu2025codexgraph,
  title={Codexgraph: Bridging large language models and code repositories via code graph databases},
  author={Liu, Xiangyan and Lan, Bo and Hu, Zhiyuan and Liu, Yang and Zhang, Zhicheng and Wang, Fei and Shieh, Michael Qizhe and Zhou, Wenmeng},
  booktitle={Proceedings of the 2025 Conference of the Nations of the Americas Chapter of the Association for Computational Linguistics: Human Language Technologies (Volume 1: Long Papers)},
  pages={142--160},
  year={2025}
}

@article{yang2025enhancing,
  title={Enhancing repository-level software repair via repository-aware knowledge graphs},
  author={Yang, Boyang and Ren, Jiadong and Jin, Shunfu and Liu, Yang and Liu, Feng and Le, Bach and Tian, Haoye},
  journal={arXiv preprint arXiv:2503.21710},
  year={2025}
}

@article{yu2025orcaloca,
  title={Orcaloca: An llm agent framework for software issue localization},
  author={Yu, Zhongming and Zhang, Hejia and Zhao, Yujie and Huang, Hanxian and Yao, Matrix and Ding, Ke and Zhao, Jishen},
  journal={arXiv preprint arXiv:2502.00350},
  year={2025}
}

@article{antoniades2024swe,
  title={Swe-search: Enhancing software agents with monte carlo tree search and iterative refinement},
  author={Antoniades, Antonis and {\"O}rwall, Albert and Zhang, Kexun and Xie, Yuxi and Goyal, Anirudh and Wang, William},
  journal={arXiv preprint arXiv:2410.20285},
  year={2024}
}

@article{chen2025unveiling,
  title={Unveiling Pitfalls: Understanding Why AI-driven Code Agents Fail at GitHub Issue Resolution},
  author={Chen, Zhi and Ma, Wei and Jiang, Lingxiao},
  journal={arXiv preprint arXiv:2503.12374},
  year={2025}
}

@article{zhang2024diversity,
  title={Diversity empowers intelligence: Integrating expertise of software engineering agents},
  author={Zhang, Kexun and Yao, Weiran and Liu, Zuxin and Feng, Yihao and Liu, Zhiwei and Murthy, Rithesh and Lan, Tian and Li, Lei and Lou, Renze and Xu, Jiacheng and others},
  journal={arXiv preprint arXiv:2408.07060},
  year={2024}
}

@article{zhang2025qwen3,
  title={Qwen3 Embedding: Advancing Text Embedding and Reranking Through Foundation Models},
  author={Zhang, Yanzhao and Li, Mingxin and Long, Dingkun and Zhang, Xin and Lin, Huan and Yang, Baosong and Xie, Pengjun and Yang, An and Liu, Dayiheng and Lin, Junyang and others},
  journal={arXiv preprint arXiv:2506.05176},
  year={2025}
}

\end{document}